\documentclass{article}

\usepackage{arxiv}

\usepackage[utf8]{inputenc} 
\usepackage[T1]{fontenc}    
\usepackage{hyperref}       
\usepackage{url}            
\usepackage{booktabs}       
\usepackage{amsfonts}       
\usepackage{nicefrac}       
\usepackage{microtype}      
\usepackage{lipsum}		
\usepackage{graphicx}
\usepackage{doi}

\usepackage{amssymb}
\usepackage{amsmath}
\usepackage{txfonts}
\usepackage{mathdots}
\usepackage[classicReIm]{kpfonts}
\usepackage{graphicx}
\usepackage{xcolor}
\usepackage{setspace}
\usepackage{times}
\usepackage{fancyhdr} 
\usepackage{scrextend}
\usepackage{hyperref}
\usepackage{enumitem}
\usepackage{indentfirst}
\usepackage{pifont}
\usepackage{amsmath,amssymb,amsfonts,wasysym}
\usepackage{algorithmic}
\usepackage{graphicx}
\usepackage{textcomp}
\usepackage{xcolor}
\usepackage{float} 
\usepackage{multirow}
\usepackage{rotating}
\usepackage{array}
\usepackage{pdflscape}
\usepackage{todonotes}
\usepackage[most]{tcolorbox}
\usepackage{xcolor} 
\usepackage{tikz}
\usepackage{subcaption}
\usepackage{multirow}
\usepackage{booktabs}
\usepackage{array}
\usepackage{tikz}
\usepackage{listings} 
\usepackage{colortbl}
\usepackage{stfloats}  
\usepackage{booktabs}
\usepackage{multirow}
\usepackage{graphicx}
\usepackage{adjustbox}

\title{A Global Analysis of Cyber Threats to \\the Energy Sector: “Currents of Conflict” \\from a geopolitical perspective}

\author{
  \href{https://orcid.org/0000-0001-6634-8315}{\includegraphics[scale=0.06]{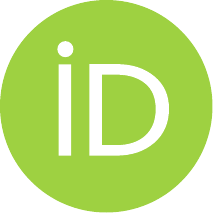}\hspace{1mm}Gustavo Sánchez}\thanks{Corresponding author.} \\
  KASTEL Security Research Labs \\
  Karlsruhe Institute of Technology (KIT) \\
  Karlsruhe, Germany \\
  \texttt{sanchez@kit.edu} \\
  \And
  \href{https://orcid.org/0000-0003-1137-1782}{\includegraphics[scale=0.06]{orcid.pdf}\hspace{1mm}Ghada Elbez} \\
  KASTEL Security Research Labs \\
  Karlsruhe Institute of Technology (KIT) \\
  Karlsruhe, Germany \\
  \texttt{ghada.elbez@kit.edu} \\
  \And
  \href{https://orcid.org/0000-0002-3572-9083}{\includegraphics[scale=0.06]{orcid.pdf}\hspace{1mm}Veit Hagenmeyer} \\
  KASTEL Security Research Labs \\
  Karlsruhe Institute of Technology (KIT) \\
  Karlsruhe, Germany \\
  \texttt{veit.hagenmeyer@kit.edu} \\
}

 \renewcommand{\headeright}{A Postprint}
\renewcommand{\undertitle}{This is a postprint of a peer-reviewed article, please cite it if using this work:\\\textit{Gustavo Sanchez, Ghada Elbez, and Veit Hagenmeyer. "A Global Analysis of Cyber Threats to the Energy Sector:“Currents of Conflict” from a geopolitical perspective." atp magazin 67.9 (2025): 56-66. \url{https://doi.org/10.17560/atp.v67i9.2797}}}
\renewcommand{\shorttitle}{A Global Analysis of Cyber Threats to the Energy Sector}

\hypersetup{
pdftitle={A template for the arxiv style},
pdfsubject={q-bio.NC, q-bio.QM},
pdfauthor={David S.~Hippocampus, Elias D.~Striatum},
pdfkeywords={First keyword, Second keyword, More},
}

\begin{document}
\maketitle

\begin{abstract}
	The escalating frequency and sophistication of cyber threats increased the need for their comprehensive understanding. This paper explores the intersection of geopolitical dynamics, cyber threat intelligence analysis, and advanced detection technologies, with a focus on the energy domain. We leverage generative artificial intelligence to extract and structure information from raw cyber threat descriptions, enabling enhanced analysis. By conducting a geopolitical comparison of threat actor origins and target regions across multiple databases, we provide insights into trends within the general threat landscape. Additionally, we evaluate the effectiveness of cybersecurity tools ---with particular emphasis on learning-based techniques--- in detecting indicators of compromise for energy-targeted attacks. This analysis yields new insights, providing actionable information to researchers, policy makers, and cybersecurity professionals.
\end{abstract}


\section{Introduction}\label{sec:Introduction}
In the evolving landscape of cybersecurity, geopolitics plays a defining role in shaping cyber threats, attack motivations, and the strategies employed by both state and non-state actors~\cite{nocetti2018geopolitics}. The emergence of regional hotspots has led to the formation of cyber alliances, where nations align their cyber defense and offensive capabilities with strategic partners. Countries within alliances such as NATO and BRICS often engage in coordinated cyber operations~\cite{smeets2019nato,belli2021cybersecurity,fracalossi2020whither}. 
Target selection in cyber conflicts is deeply rooted in geopolitical tensions, with adversarial states strategically targeting government agencies, critical infrastructure, and economic assets of rival nations. Thus, scientific literature~\cite{leventopoulos2024malware} has identified the role of malware as a pivotal geopolitical tool in the twenty-first century’s ever-evolving landscape of international relations and cybersecurity. 
Cyberattacks in general are often used as instruments of power projection, deterrence, and asymmetric warfare~\cite{todd2009armed}, allowing smaller nations or non-state actors to challenge more technologically advanced adversaries.

However, there is a lack of systematic analysis to verify this intuition across different sources. Gathering insights from different sources is important to avoid potential bias. 
Beyond direct cyber conflicts, geopolitical factors also shape national cybersecurity practices. Governments have increasingly adopted cyber resilience policies, enforcing stricter regulations on digital infrastructure and cybersecurity cooperation with allies~\cite{fracalossi2020whither}. Meanwhile, adversarial states have developed sophisticated cyber espionage and disruption campaigns, often backed by state-sponsored threat actors~\cite{hunter2021factors}. 
In this regard, Artificial Intelligence (AI) components are now a well-established part of the cyber toolbox, both for defensive and offensive purposes~\cite{truong2020artificial}.
One of the most vulnerable sectors to geopolitical cyber threats is the energy industry~\cite{maliarchuk2019hybrid,pollard2024case,whitehead2017ukraine}. As nations seek to control global energy markets, cyberattacks against energy infrastructure have become a critical component of geopolitical maneuvering. 
 
The increasing reliance on digital control systems within the energy sector~\cite{baidya2021reviewing} has made it a prime target for cyber-physical attacks, raising concerns over the resilience of national energy security in the face of rising geopolitical tensions.
However, existing databases and analytical frameworks often fall short in effectively capturing and analyzing the geopolitical nuances and sectoral focus of cyber threats.
This paper aims to address these challenges through three key contributions: 
\begin{itemize}

\item We leverage Generative AI to extract, structure and interpret information from raw cyber threat descriptions for analytical purposes.

\item We conduct a geopolitical analysis with emphasis on the origins and target regions of threat actors and cyber incidents, comparing general trends across databases with those specific to the energy sector.

\item We assess the effectiveness of firewalls in detecting Indicators of Compromise (IOCs) for attacks targeting the energy sector, with a focus on AI-based detection.

\end{itemize}

Additionally, we share the data and code used for this study with the research community\footnote{https://github.com/gus5298/SecurityThreatsGeopolitics}. 

\begin{table}[h]
\scriptsize
\centering
\caption{Databases used in this work.}
\label{tab:databases}
\resizebox{\columnwidth}{!}{%
\begin{tabular}{>{\centering\arraybackslash}p{1cm}  >{\centering\arraybackslash}p{2cm}  >{\centering\arraybackslash}p{1cm}  >{\centering\arraybackslash}p{1.6cm}  >{\centering\arraybackslash}p{5.5cm}  >{\centering\arraybackslash}p{4.5cm} }
\toprule
\rowcolor{gray!10}
\textbf{Type} & \textbf{Database} & \textbf{Country} & \textbf{Samples} & \textbf{Geographical origins and target regions} & \textbf{Target sectors} \\
\midrule
\multirow{3}{*}{Actors} 
  & MITRE ATT\&CK  & USA     & 163   & Often reported in group description text & Often reported in group description text \\ \cmidrule(lr){2-6}
  & ThaiCERT             & Thailand& 499   & Reported explicitly in \textit{JSON} format & Often reported in group description text \\  \cmidrule(lr){2-6}
  & Malpedia             & Germany & 763   & Origin reported explicitly in \textit{JSON} format, target sometimes reported in group description text & Often reported in group description text \\ 
\midrule
\multirow{2}{*}{Incidents} 
  & EuRepoC              & Germany & 3329  & Reported explicitly in tabular form & Reported explicitly in tabular form \\ \cmidrule(lr){2-6}
  & CSIS                 & USA     & 580   & Reported in incident description text & Reported in incident description text \\
\midrule
\multirow{1}{*}{Reports} 
  & AIID                 & USA     & 825     & Sometimes reported within description & Sometimes reported within description \\
\midrule
\multirow{1}{*}{Malware} 
  & Malpedia +VirusTotal API & Germany & 3166 families +2400 IOCs   & Sometimes reported, e.g., via external URLs & Sometimes reported, e.g., via external URLs \\
\bottomrule
\end{tabular}%
}
\end{table}

\section{Background and Related Work}

\textbf{Threat Actors in the Energy Domain.}
The energy sector faces a unique blend of cyber threats stemming from a wide range of actors, including state-sponsored groups, cybercriminal organizations, and hacktivists~\cite{sande2024threat}. State-sponsored threat groups are among the most sophisticated adversaries in the energy domain, often leveraging advanced persistent threats (APTs) to conduct espionage, sabotage, or influence operations~\cite{lu2024detecting}. These groups operate with the backing of national intelligence agencies, targeting energy infrastructure to gather intelligence on resource distribution, energy policies, and technological advancements, or to disrupt an adversary’s energy supply as a form of economic warfare.
The energy sector was ranked 19th out of 25 sectors in terms of actual victims of ransomware attacks in 2023~\cite{pwc_under_the_lens}. These groups exploit vulnerabilities in operational technology (OT) and industrial control systems (ICS) to extort payments from energy companies~\cite{nguyen2024towards}, with some even selling stolen data on dark web marketplaces~\cite{pantelis2021strengthening}. 
Hacktivists and ideologically driven actors also pose a significant threat to the energy sector~\cite{sande2024threat}, often launching attacks to promote environmental causes, protest against fossil fuel reliance, or disrupt operations of multinational energy corporations. 
 Emerging trends such as AI-driven cyber threats indicate that the energy sector will remain a primary target for cyber adversaries.

\textbf{Evaluation of AI-Based Detection.}
 AI-based detection systems offer several advantages, including their ability to analyze vast amounts of data in real time, identify patterns of malicious behavior, and adapt to new attack techniques through continuous learning. These technologies have significantly improved the detection of sophisticated cyber threats, particularly in identifying zero-day vulnerabilities~\cite{ahmad2023zero}, anomalies~\cite{siniosoglou2021unified}, and multi-stage attacks~\cite{jia2023artificial} that traditional signature-based detection methods may miss.
However, despite their strengths, AI/ML-based cybersecurity solutions face several limitations and challenges~\cite{arp2022and}. One major issue is the susceptibility of AI models to adversarial attacks~\cite{sanchez2024attacking}, where threat actors manipulate input data, e.g., to evade detection~\cite{mumrez2023evasion} by an intrusion detection system (IDS). Additionally, the reliance on historical data for training AI models can lead to biases~\cite{andresini2021insomnia}, making them less effective against novel attack techniques. The interpretability of AI-driven decisions remains another challenge~\cite{neupane2022explainable}.
 While AI enhances threat detection, human analysts are still essential for contextualizing threats, investigating alerts, and making strategic response decisions. 
 The ongoing development of explainable AI (XAI)~\cite{moustafa2023explainable} and hybrid AI-human collaboration models aims to address these challenges by improving the transparency and reliability of AI-driven cybersecurity solutions.
 However, to fully realize the benefits of AI in cybersecurity, more effort is required to mitigate its limitations and enhance its adaptability.

\textbf{Related Work.}
Recent research has shed light on the multifaceted challenges of detecting and mitigating cyber threats, with some looking into topics related to geopolitics:

Yuan \textit{et al.}~\cite{yuan2025beyond} offer a comprehensive threefold contribution. Their measurement study assesses the effectiveness of existing Phishing Website Detection (PWD) systems across diverse regions, revealing limitations in current approaches. Building on this, they propose enhancements to adapt PWD techniques for both Western and Chinese websites, and underscore the urgency of the issue by releasing all associated tools and datasets to stimulate real-world solution development.
Skopik \textit{et al.}~\cite{skopik2024application} take a different angle by presenting a tool that categorizes news items through advanced machine learning algorithms. By extracting and indexing key entities (such as company names, products, CVEs, and attacker groups) their tool groups related news into coherent “stories”. This approach not only aids in the rapid identification of emerging trends but also automates report summarization and leverages a collaborative ranking system to prioritize critical information.
Turning to smart grid security, Sande-Ríos \textit{et al.}~\cite{sande2024threat} provide a detailed analysis of the adversaries targeting these critical infrastructures, making reference to geopolitics. They build an adversarial model that considers the attack surface, adversary motivations, goals, and capabilities. 
Regarding analysis of IOCs, Van Liebergen \textit{et al.}~\cite{van2023deep} analyze the VirusTotal file feed over one year, reviewing 328 million reports for 235 million samples. Their study reveals that despite a feed volume 17 times lower than that of antivirus telemetry, VirusTotal detects eight times more malware.

While these contributions improve our perception of cyber threat detection and response, several research gaps remain. First, comparing diverse data sources across geopolitical contexts—particularly within the smart grid environment— requires further exploration. Second, enhancing comprehension of AI's contribution and performance metrics is necessary. Lastly, despite advancements in automating categorization and summarization, developing systems capable of integrating heterogeneous data sources into cohesive, actionable insights remains a critical requirement. In this paper, we address these gaps.

\section{Parsing Methodology}

\subsection{Databases}

We identify relevant databases from varied origins and in different formats. We introduce these sources in this subsection; an overview can be found in Table~\ref{tab:databases}.

\textbf{(1) MITRE ATT\&CK~\cite{mitre_attack_groups}.}
A globally recognized framework created by the Mitre Corporation (United States) that documents known adversary tactics, techniques and procedures. Their ATT\&CK Groups database focuses on cataloging APT groups and cybercriminal organizations, detailing their methods, tools, and motivations. This resource is widely used for threat intelligence and security strategy development. They also provide advanced information on malware families.
\textbf{(2) ThaiCERT APT Groups~\cite{etda_apt_groups}.}
 A database maintained by Thailand's national cybersecurity agency. It includes detailed profiles of APT groups, their campaigns, tools, and tactics, with information on their impact. This datasets overlaps with Malpedia~\cite{malpedia_actors}, so it is not investigated further.
\textbf{(3) Malpedia~\cite{malpedia_actors}.}
An open platform hosted by Fraunhofer FKIE (Germany) dedicated to providing detailed information on malware families and their associated threat groups. It offers comprehensive descriptions of threat actors, their campaigns, and the malware they deploy, facilitating cross-referencing between malware and the groups behind them.
\textbf{(4) EuRepoC~\cite{eurepoc_tableview}.}
The European Repository of Cyber Incidents (EuRepoC) is an independent research consortium established to enhance understanding of the cyber threat landscape within the European Union and globally. Its primary mission is to promote data-driven discussions and policy making in cybersecurity while raising awareness of cyber threats. EuRepoC provides an analytical framework to assess and compare the lifecycle of cyber incidents, with a focus on technical, political, and legal dimensions.
\textbf{(5) CSIS Signifcant Cyber Events List~\cite{csis_significant_cyber_incidents}.}
Based in the United States, the Center for Strategic and International Studies (CSIS) think tank maintains a Significant Cyber Events database, tracking major cybersecurity events worldwide. 
\textbf{(6) AIID reports~\cite{incidentdatabase_incidents}.}
The Artificial Intelligence Incident Database (AIID) is a centralized repository that documents real-world incidents involving artificial intelligence systems. It includes cases of AI-related vulnerabilities, misuse, and failures in various domains.
\newline
\textbf{Note.} The Chinese CERT~\cite{certorgcn55} releases geographical information in monthly textual summaries, but only distinguishes between “national” and “cross-border” incidents, limiting granularity and therefore is not considered in this study.

\subsection{Data Processing with Generative AI}

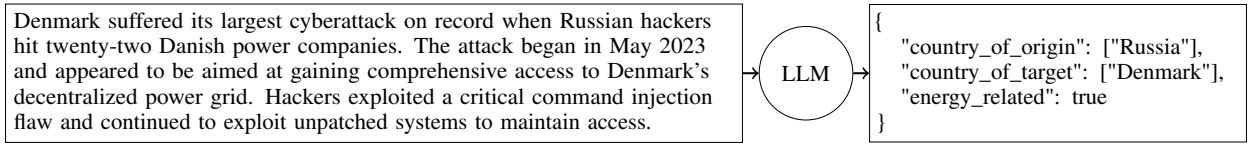
\begin{figure}[htbp]
\centering
\resizebox{\columnwidth}{!}{%
  \begin{tikzpicture}[font=\small, node distance=2.5cm]
    \node[draw, rectangle, align=left, text width=10cm] (leftbox) {
        Denmark suffered its largest cyberattack on record when Russian hackers hit twenty-two Danish power companies.
        The attack began in May 2023 and appeared to be aimed at gaining comprehensive access to Denmark's decentralized power grid.
        Hackers exploited a critical command injection flaw and continued to exploit unpatched systems to maintain access.
    };

    \node[draw, circle, minimum width=1.3cm, align=center, right of=leftbox, xshift=3.5cm] (middle) {LLM};

    \node[draw, rectangle, align=left, text width=5cm, right of=middle, xshift=1cm] (rightbox) {
    \{\\
    \quad "country\_of\_origin": ["Russia"],\\
    \quad "country\_of\_target": ["Denmark"],\\
    \quad "energy\_related": true\\
    \}
    };

    \draw[->, thick] (leftbox.east) -- (middle.west);
    \draw[->, thick] (middle.east) -- (rightbox.west);
  \end{tikzpicture}%
}
\caption{Example of a cyberattack incident description and extracted fields.}
\label{fig:cyberattack_example}
\end{figure}

Parsing unstructured threat intelligence data requires a robust and scalable approach to extract meaningful insights. In this paper, we utilize a Generative AI Model (gemini-1.5-flash-latest) to structure raw cyber incidents text data into structured fields for further analysis, as exemplified in Figure~\ref{fig:cyberattack_example}. The parsing strategy is designed to minimize Large Language Model (LLM) token consumption.

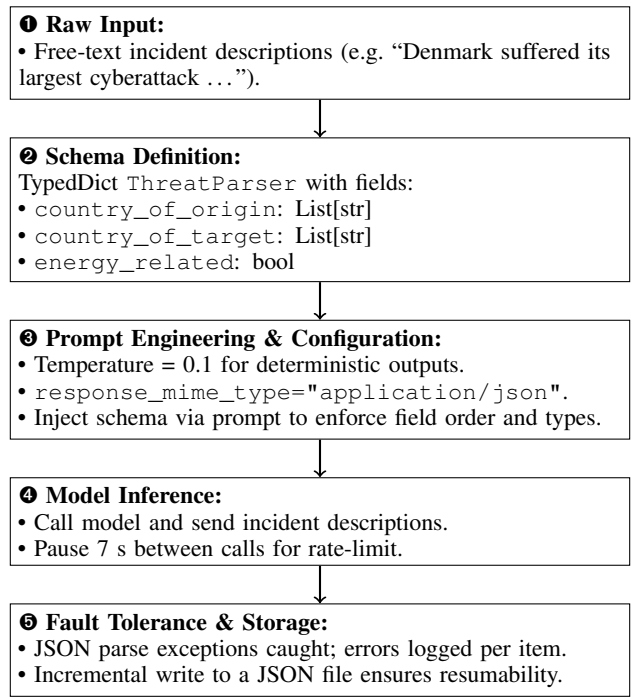
\begin{figure}[htbp]
  \centering
  \begin{tikzpicture}[font=\small, node distance=0.5cm, auto]
    \node[draw, rectangle, align=left, text width=8cm] (raw) {
      \textbf{\ding{202} Raw Input:}\\
      • Free‐text incident descriptions (e.g.\ “Denmark suffered its largest cyberattack …”).
    };
    \node[draw, rectangle, align=left, text width=8cm, below=of raw] (schema) {
      \textbf{\ding{203} Schema Definition:}\\
      TypedDict \texttt{ThreatParser} with fields:\\
      • \texttt{country\_of\_origin}: List[str]\\
      • \texttt{country\_of\_target}: List[str]\\
      • \texttt{energy\_related}: bool
    };
    \node[draw, rectangle, align=left, text width=8cm, below=of schema] (prompt) {
      \textbf{\ding{204} Prompt Engineering \& Configuration:}\\
      
      • Temperature = 0.1 for deterministic outputs.
      
• \texttt{response\_mime\_type="application/json"}.

 • Inject schema via prompt to enforce field order and types.

    };
    \node[draw, rectangle, align=left, text width=8cm, below=of prompt] (model) {
      \textbf{\ding{205} Model Inference:}\\
      • Call model and send incident descriptions. 
      
      • Pause 7 s between calls for rate‐limit.
    };
    \node[draw, rectangle, align=left, text width=8cm, below=of model] (post) {
      \textbf{\ding{206} Fault Tolerance \& Storage:}\\
       • JSON parse exceptions caught; errors logged per item.

• Incremental write to a JSON file ensures resumability.

    };

    \draw[->, thick] (raw) -- (schema);
    \draw[->, thick] (schema) -- (prompt);
    \draw[->, thick] (prompt) -- (model);
    \draw[->, thick] (model) -- (post);
  \end{tikzpicture}
  \caption{Flow of the generative‐AI parsing pipeline, from raw description to structured JSON.}
  \label{fig:parser_flowchart}
\end{figure}


As represented in Figure~\ref{fig:parser_flowchart}, the parsing workflow begins with loading the input dataset containing unstructured threat intelligence data (\ding{202}). 
The next step is codifying our target JSON format in a Python \textit{TypedDict}\footnote{A TypedDict type represents dictionary objects with a specific set of string keys, and with specific value types for each valid key.}, that we name \textit{ThreatParser} (\ding{203}). Embedding this schema directly in the system prompt ensures that Gemini outputs precisely the expected keys (and in the correct order), avoiding downstream validation errors.
We empirically tuned the model’s temperature to 0.1—balancing variability and determinism—and enforced the target JSON format (\ding{204}).

To respect API rate limits and manage token costs, a fixed 7-seconds delay is inserted between calls (\ding{205}). All responses undergo a `try/except` JSON parse: failures (e.g., reaching the API quota limit) are recorded with an error flag (\ding{206}). Partial results are written immediately to a JSON file so that parsing can resume after any interruption. The structured output captures key features, including the attack's origin, target, and domain (i.e., whether it is energy-related or not). Finally, all parsed data is stored in JSON format, preserving structured insights for downstream research and visualization.

\subsection{Evaluation}

To assess the performance of our generative‐AI parser in classifying descriptions as energy‐related, we created a stratified evaluation set of 200 threat descriptions (100 energy, 100 non‐energy) drawn at random from the EuRepoC dataset. Each entry was labeled based on the \texttt{receiver category} attribute,  and then fed to the same pipeline.
The evaluation results yielded an overall accuracy of 84.0\%. 
The following confusion matrix (rows = true class; columns = predicted) provides further details:
$
\begin{pmatrix}
91 & 9 \\ 
23 & 77
\end{pmatrix}
$.
  We observe that 91 true non‐energy incidents are correctly classified, 9 non‐energy are mislabeled as energy, 23 energy are mislabeled as non‐energy, and 77 true energy are correctly identified.
From this matrix we compute additional class‐specific metrics for the energy class:

$
\text{Precision}_{\rm energy}
= \frac{77}{77 + 9} \times 100\%
\approx 89.5\%, 
\quad
\text{Recall}_{\rm energy}
= \frac{77}{77 + 23} \times 100\%
= 77.0\%, 
\quad\\
F_{1,\rm energy}
= 2 \times \frac{\text{Precision}\times \text{Recall}}{\text{Precision} + \text{Recall}}
\approx 82.7\%.
$

Upon inspection, the sentences that are not correctly labeled are (1) those that lack explicit details\footnote{An example of this is: “Likely Iranian State-sponsored hackers (Crowd strike) have conducted a series of destructive attacks on Saudi Arabia over the last two weeks, erasing data and wreaking havoc in the computerbanks of the agency running the country’s airports and hitting five additional targets”.} about the energy targets, and (2) those labeled as “Government”, e.g., in the cases where the target is the US Department of Energy. 
These results demonstrate that our schema‑enforced, low‐temperature prompting approach achieves strong overall accuracy, high precision on the energy class, and acceptable recall given the challenging, unstructured nature of threat descriptions.  

For comparison, the spaCy~\cite{spacy-phrasematcher} rule‑based baseline (using a \texttt{PhraseMatcher} with 16 domain‑specific terms\footnote{Requires manually gathering keywords via expert knowledge, while our Gen-AI approach only requires to specify the domain itself, e.g.: energy.} such as “energy”, “power grid”, etc.) reached an overall accuracy of 81\%. Its confusion matrix 
$
\begin{pmatrix}
95 & 5 \\ 
34 & 66
\end{pmatrix},
$
 corresponds to a precision of 93\% and a recall of 66\% for the energy class, giving an F\textsubscript{1} score of 77\%. 

In summary, these results demonstrate that our schema‑enforced, low‑temperature Gen-AI pipeline not only outperforms a rule‑based baseline—boosting recall by +~11 percentage points (77\% vs. 66\%) on energy‑related cases—but also maintains comparably high precision, underscoring the practical value of generative AI for accurately extracting nuanced domain signals from unstructured threat descriptions.

\begin{table}[htbp]
  \centering
  \scriptsize
  \caption{Comparison of classification metrics for energy‑domain detection.}
  \label{tab:eval_comparison}
  \begin{tabular}{lcc}
    \toprule
    \rowcolor{gray!10}
    \textbf{Metric}            & \textbf{Our Gen‑AI Parser}       & \textbf{spaCy Baseline} \\
    \midrule
    Accuracy                   & \cellcolor{green!20}84.0\%       & 81.0\%                 \\
    Precision (energy)         & 89.5\%                           & \cellcolor{green!20}93.0\% \\
    Recall (energy)            & \cellcolor{green!20}77.0\%       & 66.0\%                 \\
    F\textsubscript{1} (energy)& \cellcolor{green!20}82.7\%       & 77.0\%                 \\
    \bottomrule
  \end{tabular}
\end{table}

\section{Geopolitical Big Data Analysis Results}
The reporting of geographical origins and target regions varies significantly across different cyber threat databases, reflecting disparities in structure, detail, and accessibility. Table~\ref{tab:databases} summarizes the geographical reporting practices of major cyber threat intelligence resources.
Some databases provide structured, machine-readable formats that enable streamlined analysis. For instance, ThaiCERT and Malpedia report geographical origins explicitly in JSON format, facilitating automation and consistency. However, target regions in Malpedia are often embedded within descriptive text, requiring additional processing via the proposed generative AI pipeline. 
EuRepoC explicitly documents both origins and targets in tabular form, offering a standardized approach allowing comparative analysis. 
In contrast, databases such as MITRE ATT\&CK Groups and CSIS rely on unstructured descriptive text to document geographical information. While rich in qualitative details, these formats necessitate our processing technique to structure the data for analysis. 
Not all databases include geographical reporting. Notably, AIID omits geographical origins and targets entirely, reducing their utility for geopolitical studies. This omission prevents understanding the spatial distribution of cyber threats. Using our generative AI pipeline, we are still able to extract whether the AI vulnerabilities are related to the energy domain.

\begin{tcolorbox}[
    colback=gray!10,                  
        colframe=gray!10,                  
    left=2mm,                         
    boxrule=0mm,                      
    borderline west={0.01mm}{0mm}{red!75!black}, 
    enhanced,                         
    sharp corners,                     
        left=1mm,                          
    right=1mm,                         
    top=1mm,                           
    bottom=1mm                        
]
\textbf{Takeaway 1.} There is heterogeneity in reporting practices among cyber threat databases. Structured reporting (e.g. JSON), enhances consistency and facilitates cross-database comparisons. The reliance on descriptive text or omission of geographical information hinders the ability to conduct comprehensive geopolitical analyses. Our generative AI pipeline allows for better analysis by parsing relevant unstructured information. 
\end{tcolorbox}

\begin{figure}[h]
  \centering
  \begin{subfigure}[b]{0.49\linewidth}
    \centering
    \includegraphics[width=0.75\linewidth]{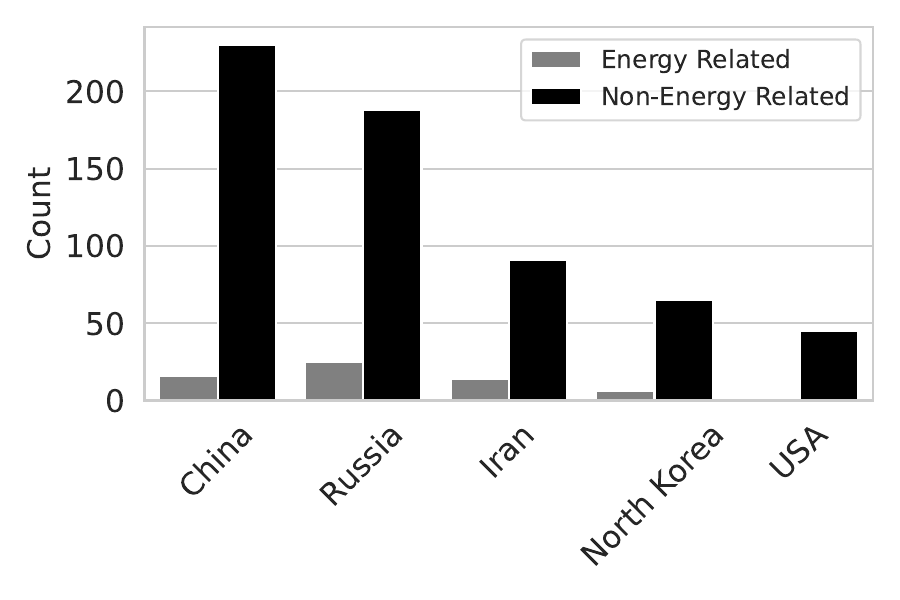}
    \caption{CSIS Origins}
    \label{fig:origins_csis}
  \end{subfigure}
  \hfill
  \begin{subfigure}[b]{0.49\linewidth}
    \centering
    \includegraphics[width=0.75\linewidth]{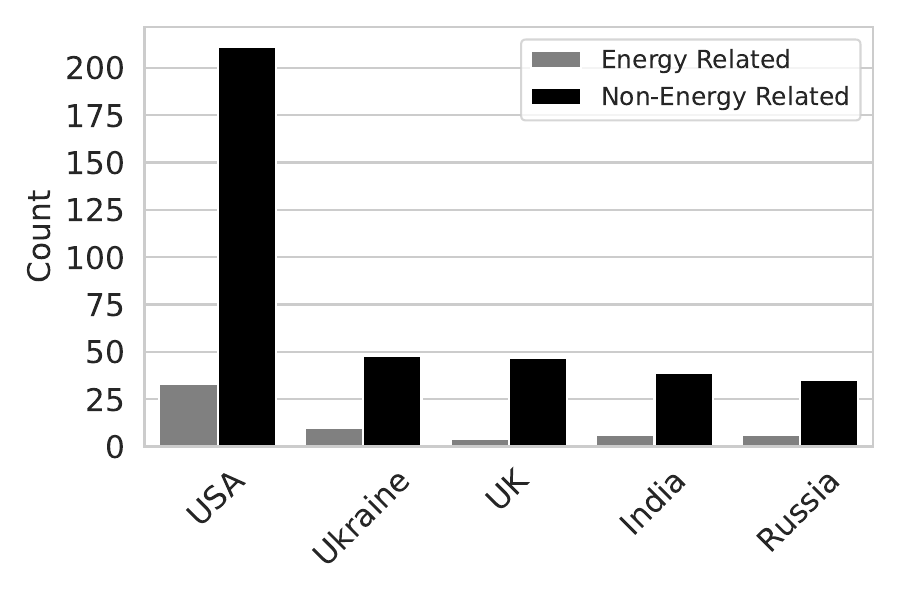}
    \caption{CSIS Targets}
    \label{fig:alliances_csis}
  \end{subfigure}
  
  
  \begin{subfigure}[b]{0.49\linewidth}
    \centering
    \includegraphics[width=0.75\linewidth]{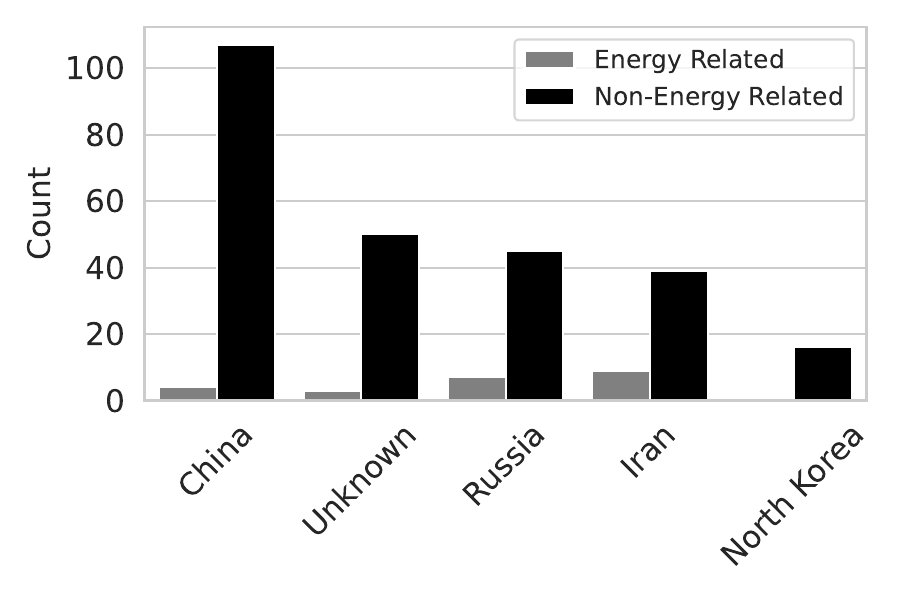}
    \caption{Malpedia Origins}
    \label{fig:origins_malpedia}
  \end{subfigure}
  \hfill
  \begin{subfigure}[b]{0.49\linewidth}
    \centering
    \includegraphics[width=0.75\linewidth]{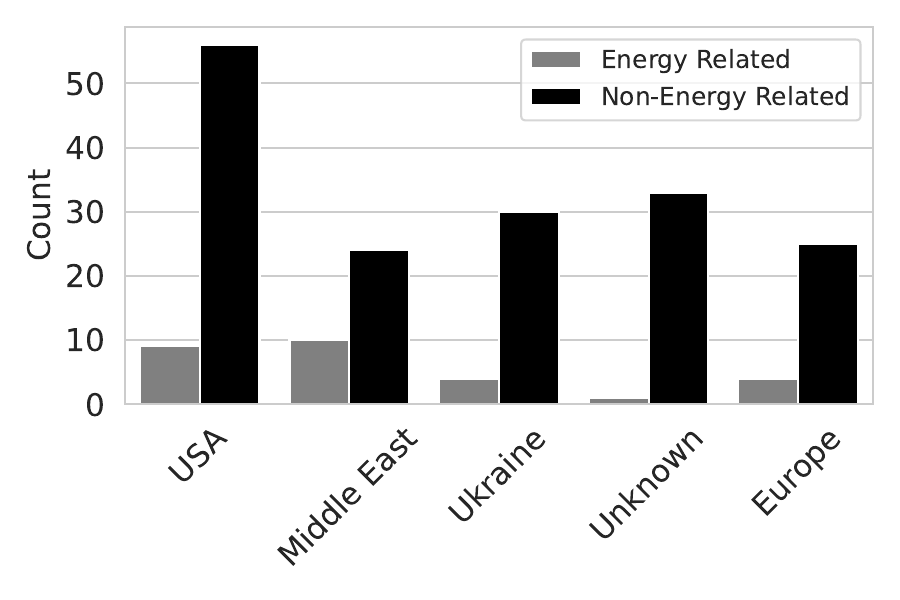}
    \caption{Malpedia Targets}
    \label{fig:alliances_malpedia}
  \end{subfigure}
  
  \begin{subfigure}[b]{0.49\linewidth}
    \centering
    \includegraphics[width=0.75\linewidth]{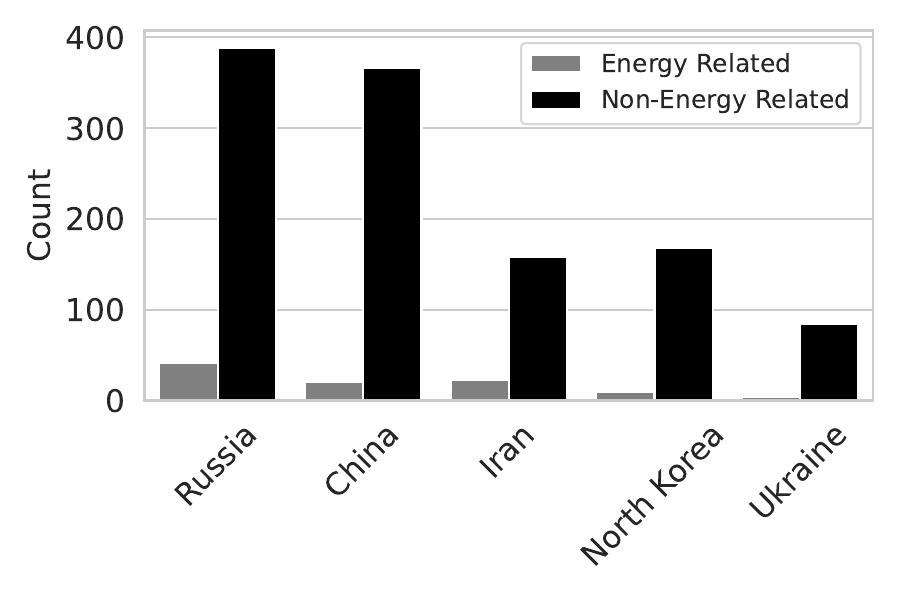}
    \caption{EuRepoC Origins}
    \label{fig:origins_eurepoc}
  \end{subfigure}
  \hfill
  \begin{subfigure}[b]{0.49\linewidth}
    \centering
    \includegraphics[width=0.75\linewidth]{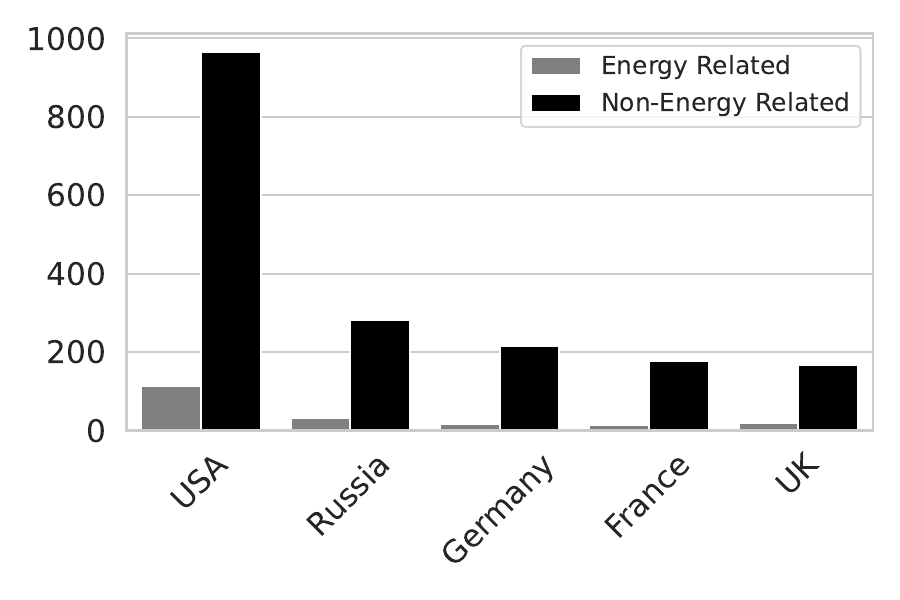}
    \caption{EuRepoC Targets}
    \label{fig:alliances_eurepoc}
  \end{subfigure}
  
  \caption{Top 5 threat origins and targets per dataset.}
  \label{fig:origins}
\end{figure}

From the provided plots in Figure~\ref{fig:origins}, each subfigure shows the top 5 of either threat origins or targets, with bars grouped by energy vs. non-energy related. The three relevant data sources to answer this question—CSIS, Malpedia, and EuRepoC—each provide a slightly different perspective on which countries appear most frequently. In general, the same few countries dominate both energy and general threats (i.e., Russia and China as origins, and the USA as target). However, the bar heights (counts) can differ substantially. For instance, a country might be the top origin overall but drop to a lower position (or even disappear) in the energy-related subset, indicating that some origins are especially active in the general domain but less so in energy.
The energy-related bars show a higher concentration in fewer countries—often, one or two countries account for most of the energy-focused incidents—suggesting that certain threat actors specialize in energy infrastructure attacks.
By contrast, the general category includes many different targets (government agencies, corporations, etc.). This can make the energy domain appear narrower but more heavily hit by a smaller set of threat actors.
The USA appears as a top general target across datasets, while in energy incidents, the Middle East as a region takes the first place by a small difference (according to Malpedia). 

\begin{figure}[h]
  \centering
  \begin{subfigure}[b]{0.49\linewidth}
    \centering
    \includegraphics[width=0.75\linewidth]{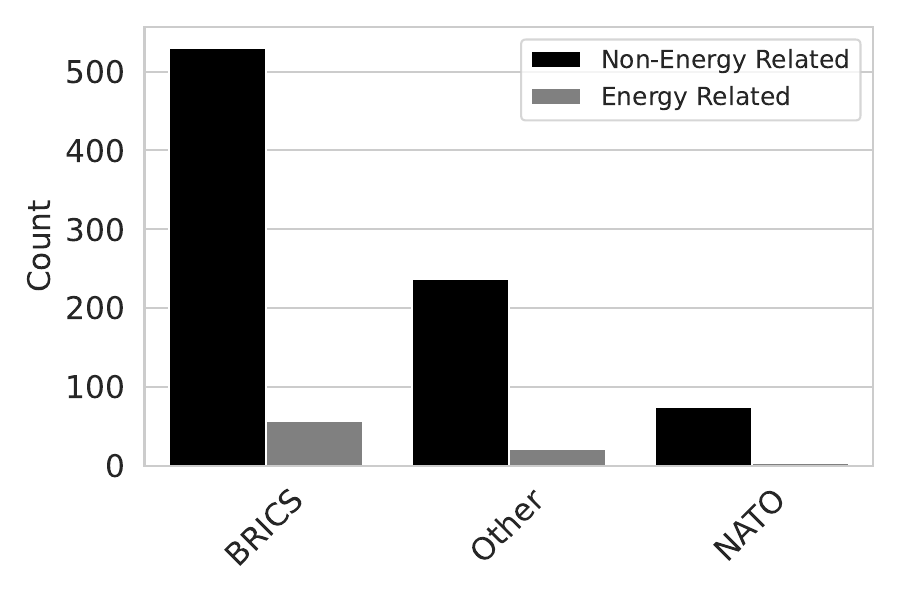}
    \caption{CSIS Origins}
    \label{fig:o_alliances_csis}
  \end{subfigure}
  \hfill
  \begin{subfigure}[b]{0.49\linewidth}
    \centering
    \includegraphics[width=0.75\linewidth]{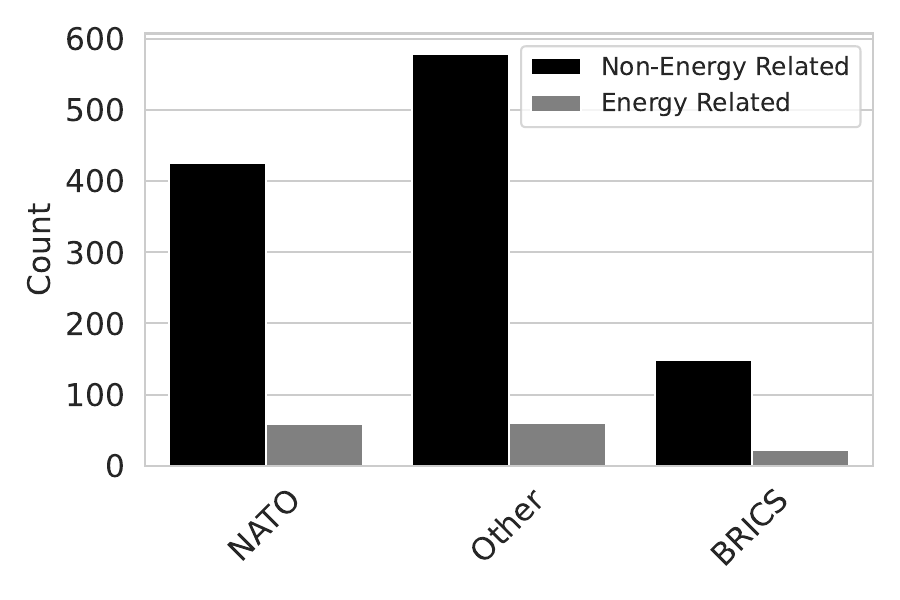}
    \caption{CSIS Targets}
    \label{fig:alliances_csis}
  \end{subfigure}
  
  
  \begin{subfigure}[b]{0.49\linewidth}
    \centering
    \includegraphics[width=0.75\linewidth]{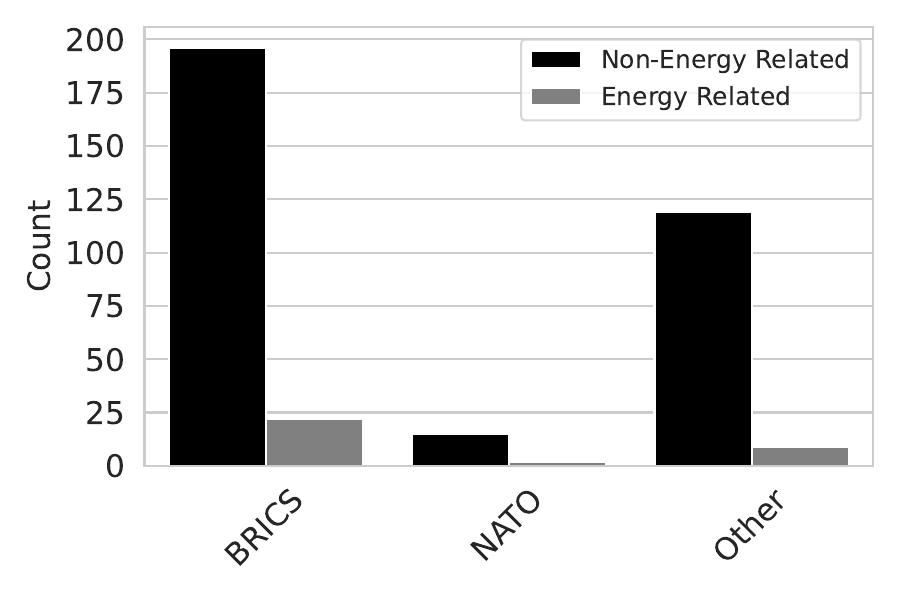}
    \caption{Malpedia Origins}
    \label{fig:alliance_origins_malpedia}
  \end{subfigure}
  \hfill
  \begin{subfigure}[b]{0.49\linewidth}
    \centering
    \includegraphics[width=0.75\linewidth]{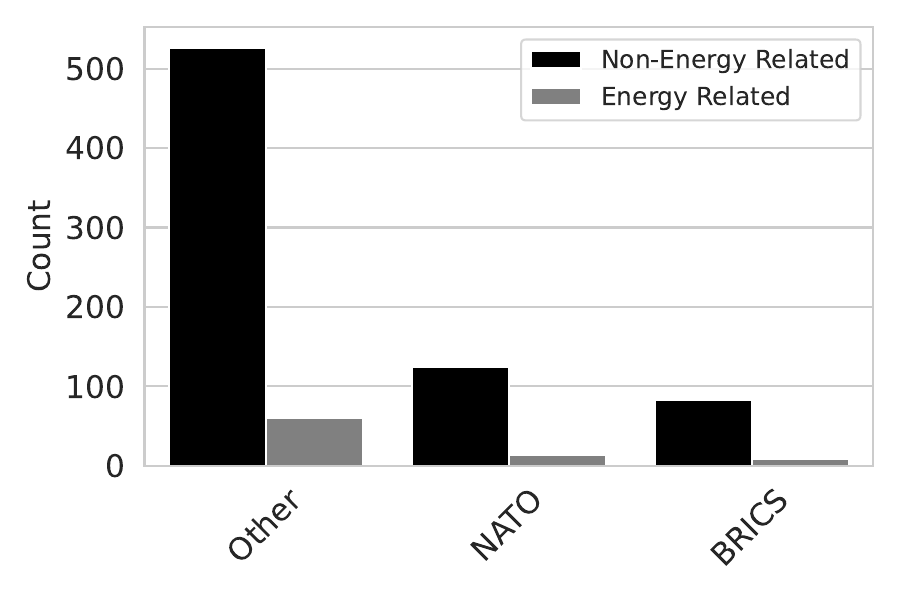}
    \caption{Malpedia Targets}
    \label{fig:alliances_target_malpedia}
  \end{subfigure}
  
  \begin{subfigure}[b]{0.49\linewidth}
    \centering
    \includegraphics[width=0.75\linewidth]{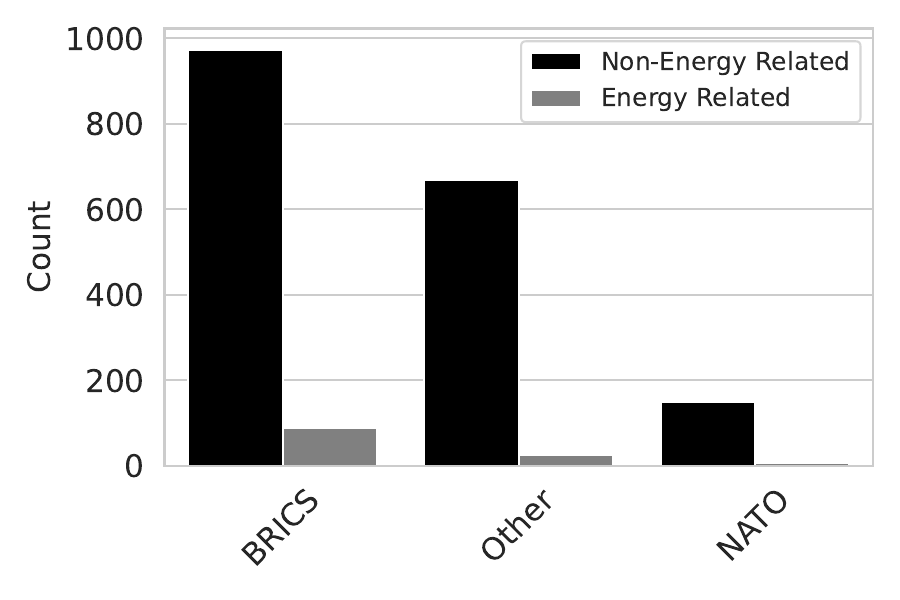}
    \caption{EuRepoC Origins}
    \label{fig:origins_eurepoc}
  \end{subfigure}
  \hfill
  \begin{subfigure}[b]{0.49\linewidth}
    \centering
    \includegraphics[width=0.75\linewidth]{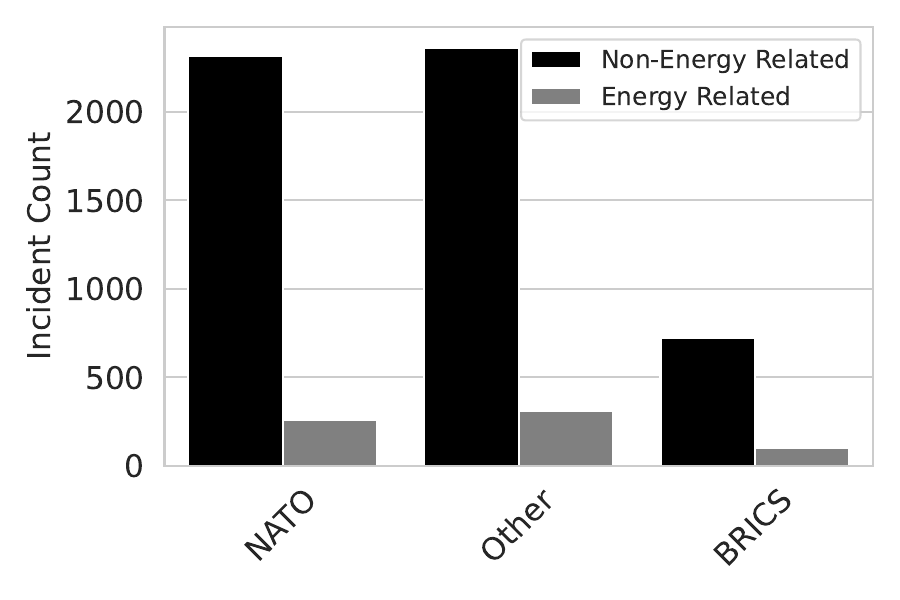}
    \caption{EuRepoC Targets}
    \label{fig:alliances_eurepoc}
  \end{subfigure}
  
  \caption{Top 5 threat origins and targets per dataset, clustered by alliances.}
  \label{fig:alliances}
\end{figure}

In Figure~\ref{fig:alliances}, we group each incident’s origin (and, analogously, its target) by major geopolitical alliances—namely NATO, BRICS, or “Other”—using up‑to‑date membership rosters as of January 2025. Specifically, our NATO list includes the 32 member states from Albania through the United States (adding Sweden, which joined in July 2023), while the BRICS bloc encompasses not only the original five (Brazil, Russia, India, China, South Africa) but also the ten countries formally participating in the 2024–25 BRICS expansion (Egypt, Ethiopia, Indonesia, Iran, and the United Arab Emirates), providing a relevant, up-to-date overview. Any country not found in either list is classified as “Other”.

Concretely, we take each record’s country-of-origin field (which may be a list), standardize the country names, and then map each to its alliance via a lookup. We then group by alliance membership; every origin country contributes separately to the aggregate counts, so that each pair (incident, country) becomes its own row. Finally, we assign an alliance label, and plot counts for “Energy Related” versus “Non‑Energy Related” in matching bar charts. This clustering by alliance reveals, for example, that BRICS countries dominate non‑energy threat origins across all datasets, whereas energy‑focused incidents show a comparatively higher concentration in “Other” or NATO members depending on the source.
In the Malpedia dataset (Figures \ref{fig:alliance_origins_malpedia} and~\ref{fig:alliances_target_malpedia}), we observe the biggest contrast between alliances.

\begin{tcolorbox}[
    colback=gray!10,                  
        colframe=gray!10,                  
    left=2mm,                         
    boxrule=0mm,                      
    borderline west={0.01mm}{0mm}{red!75!black}, 
    enhanced,                         
    sharp corners,                     
        left=1mm,                          
    right=1mm,                         
    top=1mm,                           
    bottom=1mm                        
]
\textbf{Takeaway 2.} General threats tend to involve a broader set of countries both in terms of origins and targets, whereas energy-related attacks are often more narrowly concentrated. The same few countries dominate the overall ranking—e.g., Russia, China, or the USA—but for energy-focused incidents, certain actors seem to specialize.  
\end{tcolorbox}

\begin{figure}[htbp]
  \centering
  \begin{subfigure}[b]{0.49\linewidth}
    \centering
    \includegraphics[width=\linewidth]{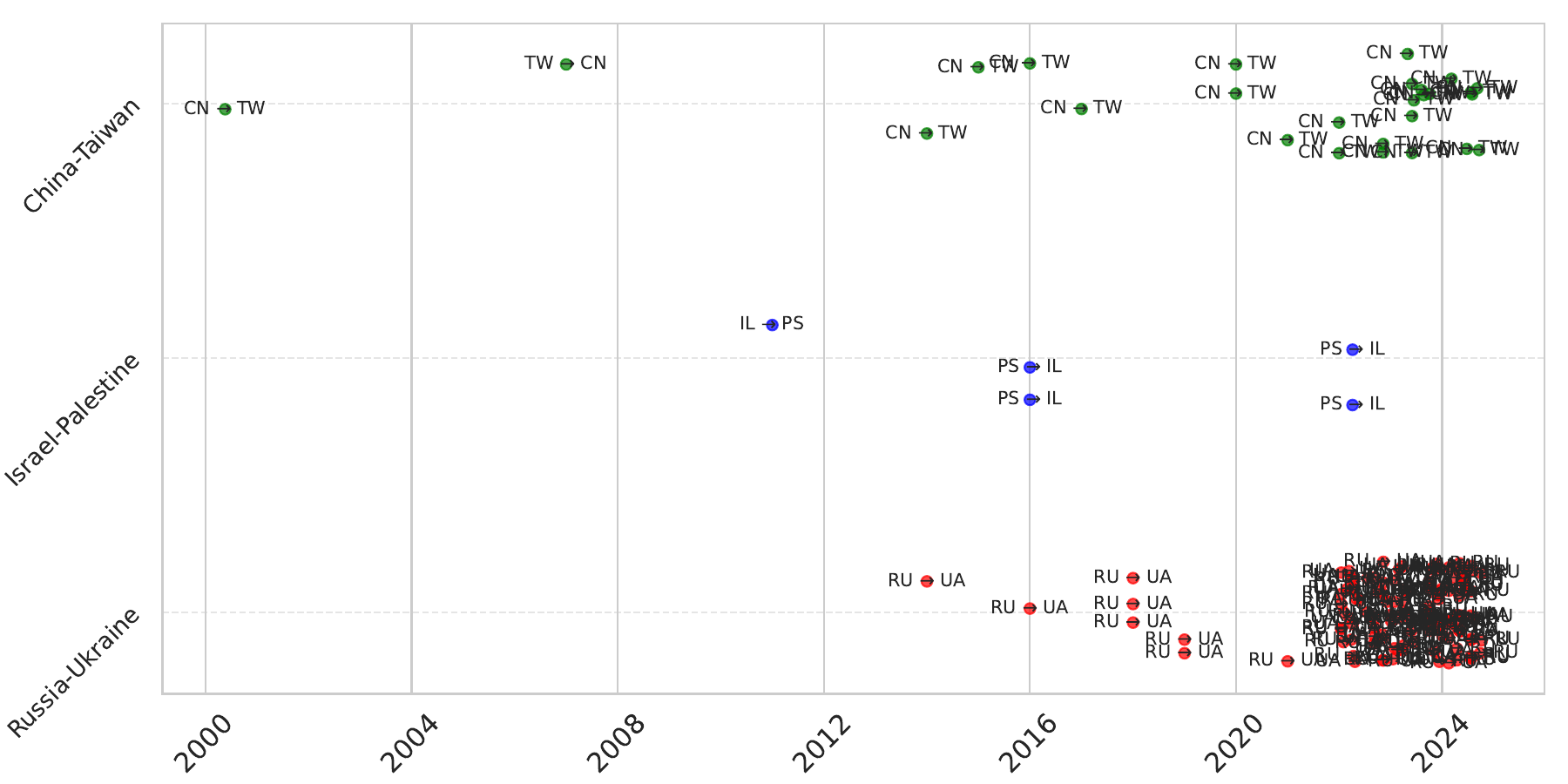}
    \caption{EuRepoC Dataset}
    \label{fig:conflics_eurepoc}
  \end{subfigure}
  \hfill
  \begin{subfigure}[b]{0.49\linewidth}
    \centering
    \includegraphics[width=\linewidth]{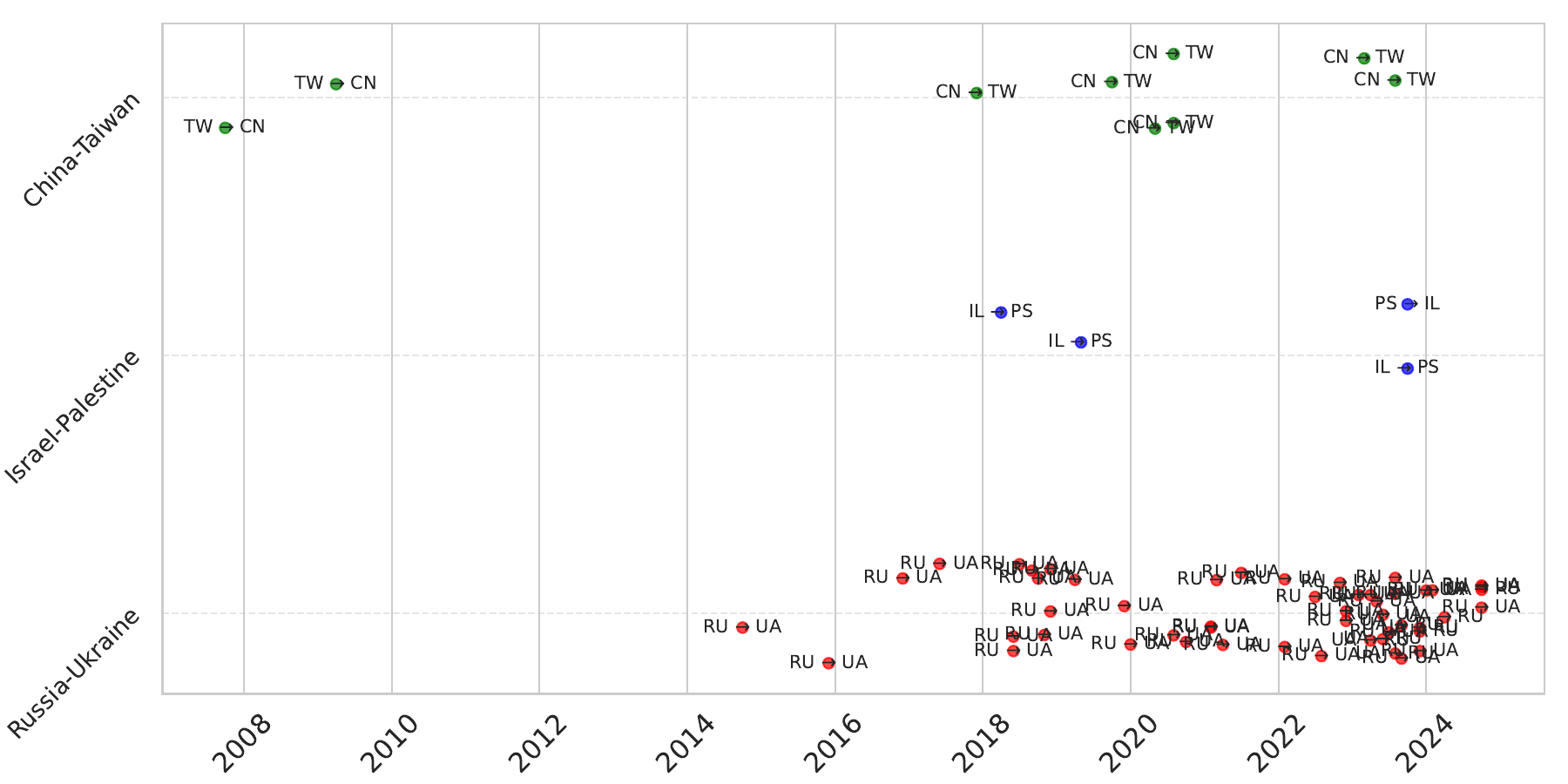}
    \caption{CSIS Dataset}
    \label{fig:conflics_csis}
  \end{subfigure}
  
  \vspace{0.5cm}
  
  \begin{subfigure}[b]{0.9\linewidth}
    \centering
    \scriptsize
    \begin{adjustbox}{max width=\linewidth}
      \begin{tabular}{c l l l l}
        \toprule
        \rowcolor{gray!10}
         & \textbf{Date} & \textbf{Initiator} & \textbf{Target} & \textbf{Details} \\
        \midrule
        \multirow{11}{*}{\rotatebox{90}{CSIS}} 
            & 2015-12 & Russia & Ukraine & DoS and SCADA attacks \\
            & 2016-12 & Russia & Ukraine & Ukrenergo shut down \\
            & 2017-06 & Russia & Ukraine & Ransomware \\
            & 2018-06 & Russia & Ukraine & Backdoors \\
            & 2020-01 & Russia & Ukraine & Infiltration \\
            & 2020-05 & China & Taiwan & Malware attacks \\
            & 2020-10 & Russia & Ukraine & Officers indicted ('16 attacks) \\
            & 2022-08 & Russia & Ukraine & State energy website hack \\
            & 2022-11 & Russia & Ukraine & Energy and logistics sector hack \\
            & 2022-12 & Russia & Ukraine & Sandworm APT \\
            & 2023-12 & Ukraine & Russia & Encryption/deletion of data \\
        \midrule
        \multirow{9}{*}{\rotatebox{90}{EuRepoC}} 
            & 2024-09-19 & China & Taiwan & Spear-phishing \\
            & 2024-04-19 & Russia & Ukraine & Sandworm APT \\
            & 2023-06-15 & China & Taiwan & CVE-2023-2868 \\
            & 2023-01-29 & Ukraine & Russia & Gazprom hack-and-leak \\
            & 2023-01-31 & Russia & Ukraine & Sandworm APT \\
            & 2022-08-17 & Russia & Ukraine & Energoatom website hack \\
            & 2022-02-28 & Russia & Ukraine & Sandworm APT \\
            & 2022-04-12 & Russia & Ukraine & Sandworm APT \\
            & 2022-04-12 & Russia & Ukraine & Sandworm APT \\
        \bottomrule
      \end{tabular}
    \end{adjustbox}
    \caption{Energy-Related Cyber Incidents in Current Conflicts}
    \label{tab:energy_conflict_incidents}
  \end{subfigure}
  
  \caption{Comparison of conflict chronologies (top row) and energy-related cyber incidents (bottom row).}
  \label{fig:combined_figure}
\end{figure}

From the plots in Figure~\ref{fig:combined_figure}, it is clear that cyber incidents in conflict regions are neither uniformly distributed over time nor concentrated in a single geographic theater. Instead, they form distinct clusters that align with specific escalations in each ongoing conflict. In the Russia–Ukraine timeline, a noticeable uptick appears around 2014 (following the annexation of Crimea) and accelerates substantially from 2022 onward. Israel–Palestine incidents show intermittent bursts—fewer overall than Russia–Ukraine, but still recurring during major flare-ups in the region. Meanwhile, the China–Taiwan timeline is comparatively sparser but displays a steady trickle of events spanning multiple years, suggesting a slow-burn pattern of cyber activity. Each conflict region thus exhibits its own rhythm of threat incidents, typically intensifying around key military escalations.

\begin{tcolorbox}[
    colback=gray!10,                  
        colframe=gray!10,                  
    left=2mm,                         
    boxrule=0mm,                      
    borderline west={0.01mm}{0mm}{red!75!black}, 
    enhanced,                         
    sharp corners,                     
        left=1mm,                          
    right=1mm,                         
    top=1mm,                           
    bottom=1mm                        
]
\textbf{Takeaway 3.} Heightened conflicts correlate with increased frequency of attacks. Russia’s persistent cyber campaigns since 2014 reflect ongoing regional disputes. In contrast, the Israel–Palestine conflict shows sharp, periodic spikes in activity that mirror its cyclical hostilities, while China–Taiwan experiences a steady, lower-intensity flow of incidents, suggesting an enduring contest of influence rather than a single, large-scale campaign. 
\end{tcolorbox}

\begin{figure}[htbp]
  \centering
  \begin{subfigure}[b]{0.49\linewidth}
    \centering
    \includegraphics[width=0.85\linewidth]{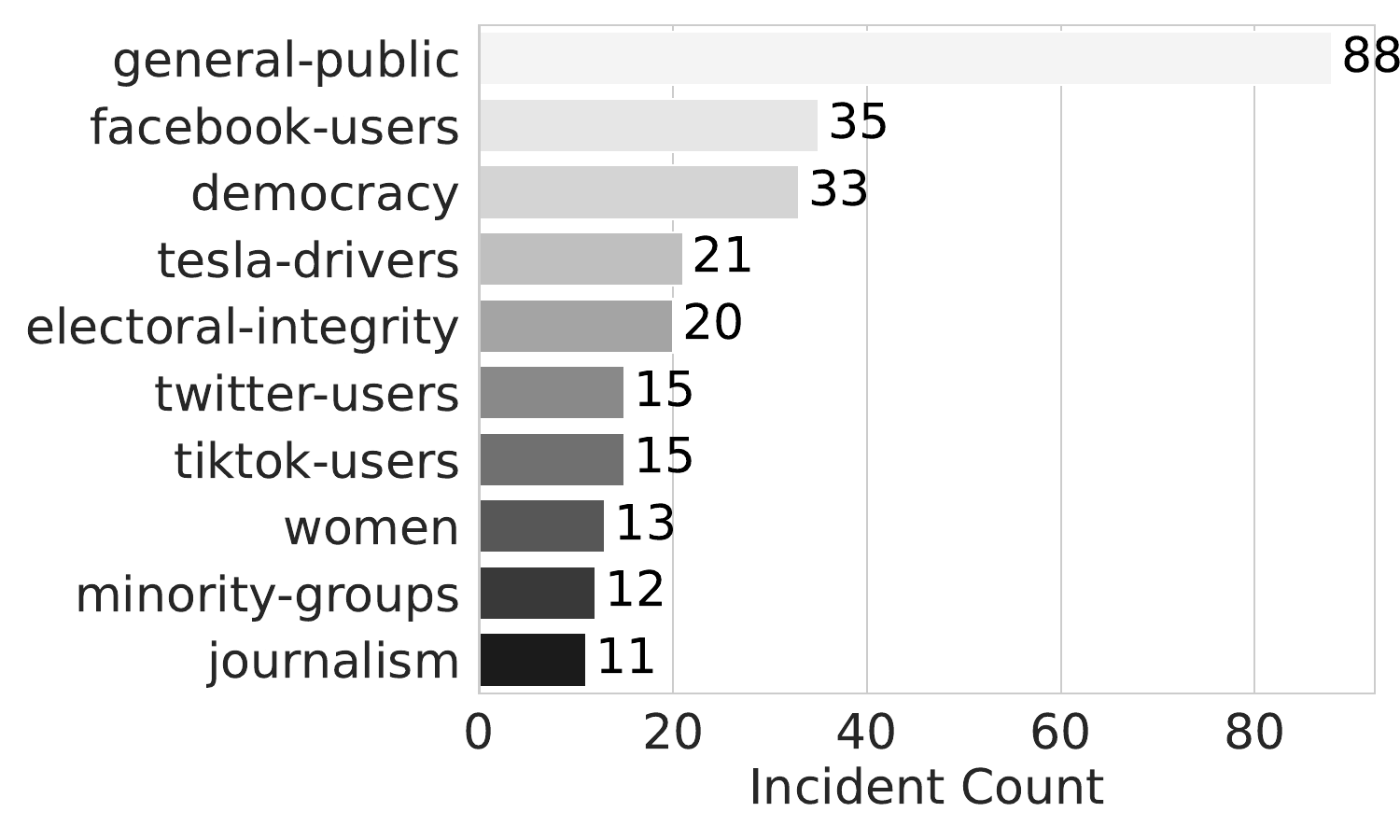}
    \caption{Top 10 Alleged Harmed Parties in the AIID dataset.}
    \label{fig:alleged_harmed}
  \end{subfigure}
  \hfill
  \begin{subfigure}[b]{0.49\linewidth}
    \centering
    \includegraphics[width=\linewidth]{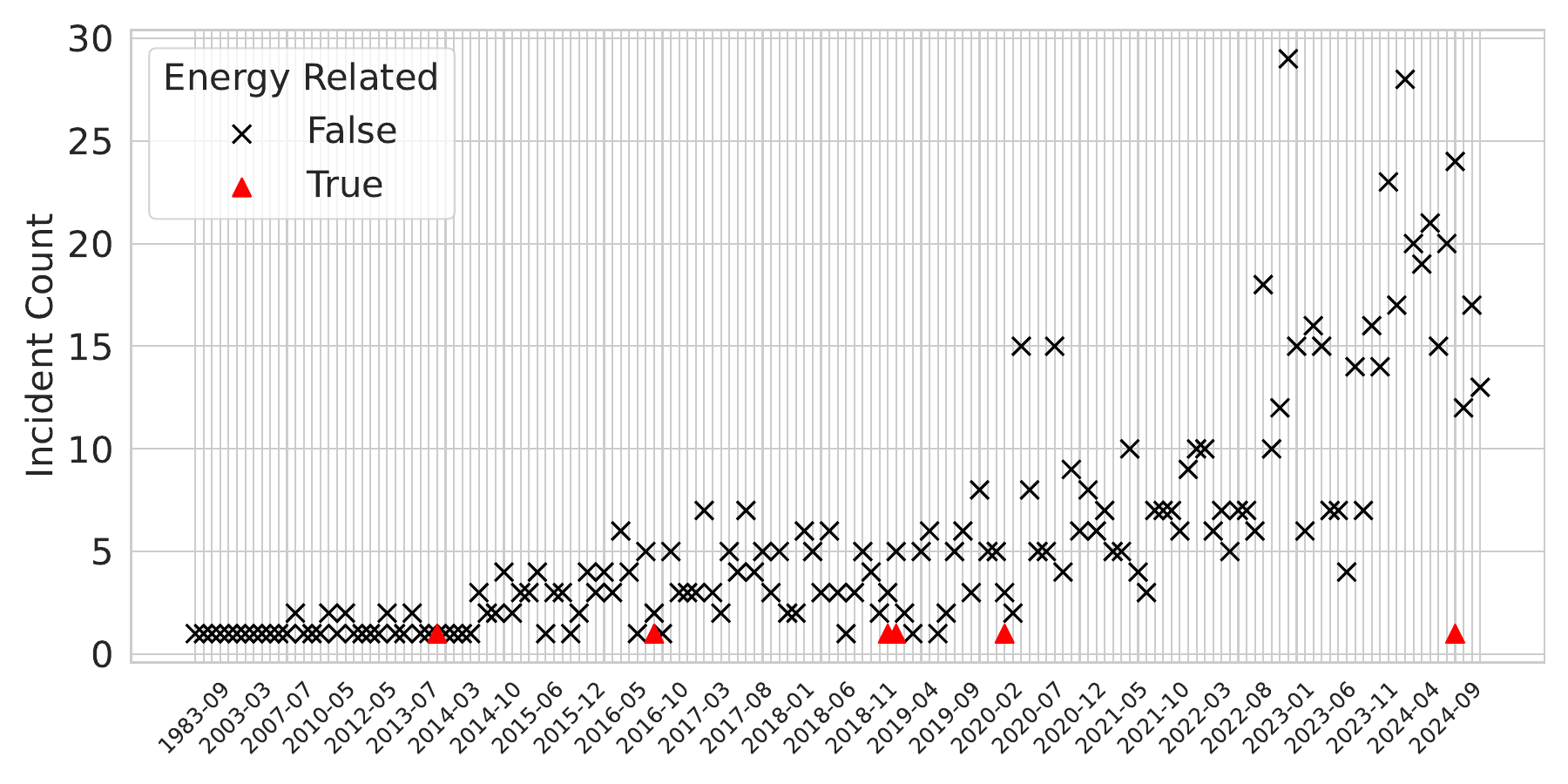}
    \caption{AI Threat Incidents Over Time in the AIID dataset.}
    \label{fig:aiid}
  \end{subfigure}
  \caption{Analysis of AI Threat Incidents in the AIID dataset.}
  \label{fig:combined_aiid}
\end{figure}

We perform a series of experiments to classify and analyze incidents from the AIID, which contains almost 800 records. The goal is to identify and categorize AI incidents related to specific topics, with a particular focus on the energy domain.
Figure~\ref{fig:alleged_harmed} shows that the top alleged harmed parties in the AIID dataset include a broad range of groups—such as general-public, democracy, or Tesla drivers—while energy- or power-sector victims do not prominently appear among the top ten. It is important to note that, in the case of Tesla cars, AI vulnerabilities lie in the autonomous driving or manufacturing automation processes, and not in the electric vehicle infrastructure. Relative to other domains, explicit AI-based incidents in the energy sector are less common or, at least, less frequently reported. Meanwhile, Figure~\ref{fig:aiid}, which plots AI threat incidents over time and highlights energy-related events with red triangles, reveals that such incidents do exist but remain comparatively sparse.  To shed light into this, we describe the specific AI and energy related incidents in Table~\ref{tab:energy_aiid_incidents}; in some of these incidents, the energy relation is relatively loose: the generative AI model considers e.g., smart home devices as part of the smart grid, and the climate crisis a subclass of the energy domain.
As AI technologies continue to integrate into critical infrastructure, including power grids and smart devices, the potential for energy-targeted attacks will likely grow.

\begin{table*}[h!]
    \centering
    \scriptsize
    \caption{Most energy-related Cyber Incidents in AIID.}
    \label{tab:energy_aiid_incidents}
    \resizebox{\columnwidth}{!}{%
    \begin{tabular}{l l l l}
        \toprule
        \rowcolor{gray!10}
          \textbf{Date} & \textbf{Alleged Developer} & \textbf{Alleged Harmed} & \textbf{Details} \\
        \midrule
        \multirow{11}{*}
        
             2016-10-08 & Tesla & Tesla & Poor Performance of Tesla Factory AI Robots (assembling lithium batteries) \\
             2014-01-21 & Nest Labs & Fire victims  & Smoke + CO Alarm could inadvertently silence genuine alarms (smart home) \\
             2019-03-01 & Scammers & UK Energy Firm's CEO & Fraudsters Used AI to Mimic Voice \\
             2020-04-14 & Belgian action group & Belgian government & Deepfake of Belgian Prime Minister Urging Climate Crisis Action \\
             2019-02-01 & YouTube & YouTube users & Recommendation Algorithm Allegedly Promoted Climate Misinformation Content \\
             2024-10-14 & Portland Water Bureau & City of Portland & Portland Water Bureau  Algorithm Reportedly Allocates Utility Bill Discount to High-Wealth Consumer \\
        \midrule
        
    \end{tabular}}
\end{table*}

\begin{tcolorbox}[
    colback=gray!10,                  
        colframe=gray!10,                  
    left=2mm,                         
    boxrule=0mm,                      
    borderline west={0.01mm}{0mm}{red!75!black}, 
    enhanced,                         
    sharp corners,                     
        left=1mm,                          
    right=1mm,                         
    top=1mm,                           
    bottom=1mm                        
]
\textbf{Takeaway 4.} These observations imply that AI-related vulnerabilities in the energy domain exist but they appear overshadowed by a wider array of AI threats affecting broader consumer or societal areas.   
\end{tcolorbox}


We analyze the performance of various antivirus engines, with an emphasis on SentinelOne and Acronis, which explicitly report the usage of static machine learning models.
We filter out cyber threat groups related to the energy domain. Then, we extract the information about tools that they leverage (e.g., malware 
 families). 
From the tool names, we fetch IOCs (i.e., file hashes) via Malpedia. 
These hashes are then submitted to the VirusTotal platform via API to gather detection results from various antivirus engines, both AI-based and traditional. 
The VirusTotal results were stored locally to avoid redundant queries, preserving API quota for future experiments.

Our analysis shows that only a fraction of IOCs (12.85\%) are explicitly tied to energy-focused threats, and that Static ML tools (Acronis and SentinelOne) detect only around 46.8\% of malicious indicators, whereas other engines (the \textit{Others} category) detect roughly 88.4\%—on par with the overall average (88.4\%). This indicates that, for the malware samples used to target the energy domain, traditional solutions are outperforming the ML-based ones.
\begin{figure*}[!htbp]
  \centering
  \includegraphics[width=1\textwidth]{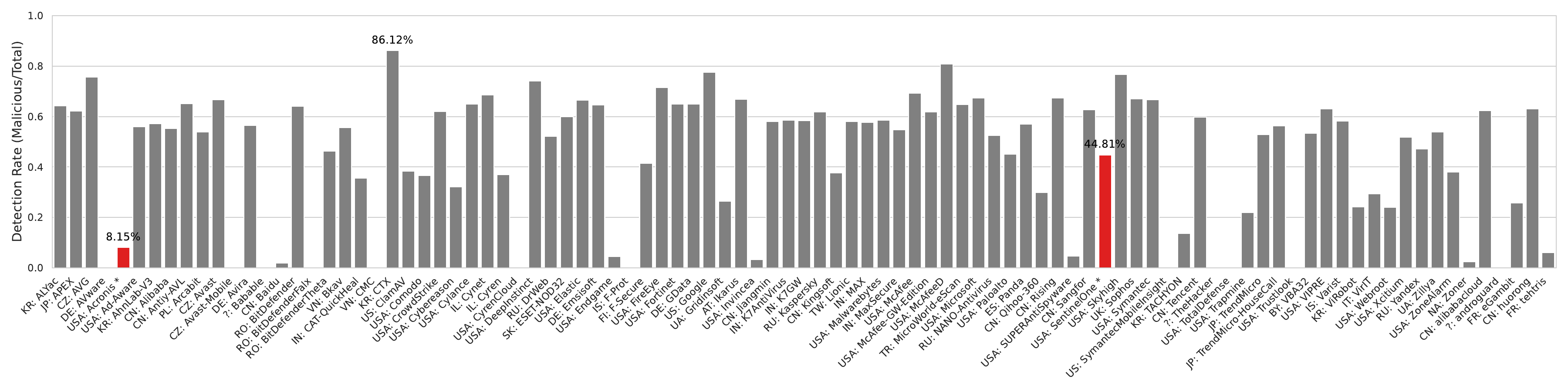}
  \caption{Detection rate by antivirus engine. Static ML engines (Acronis and SentinelOne) are highlighted in red.}
  \label{fig:detection_rate_by_engine}
\end{figure*}
Figure~\ref{fig:detection_rate_by_engine} presents a side-by-side comparison of each antivirus engine’s detection rate, with Acronis and SentinelOne (the static ML engines) emphasized in red. The top-performing engine in this dataset achieves a higher detection rate than either of these ML-based solutions. 

\begin{tcolorbox}[
    colback=gray!10,                  
        colframe=gray!10,                  
    left=2mm,                         
    boxrule=0mm,                      
    borderline west={0.01mm}{0mm}{red!75!black}, 
    enhanced,                         
    sharp corners,                     
        left=1mm,                          
    right=1mm,                         
    top=1mm,                           
    bottom=1mm                        
]
\textbf{Takeaway 5.} Although ML-based cybersecurity solutions are frequently touted as being especially agile at catching new or sophisticated attacks, the data here suggests they may lag behind more traditional or hybrid approaches when confronting malware associated with the energy domain. 
\end{tcolorbox}


In this work, we analyze the use of AI—specifically, static machine learning—as a defensive method for detecting IOCs, and generative AI to support natural language tasks. However, the malicious use of AI to enhance cyber operations is an emerging threat (unless done ethically, e.g., for web application penetration testing~\cite{sanchez2024web}) with significant geopolitical implications and the potential to impact energy systems.
Recent reports~\cite{openai2024influence} have highlighted three cases of AI-related cyber operations involving OpenAI’s ChatGPT:
 \textbf{(1) SweetSpecter}: This suspected China-based adversary uses OpenAI’s services for reconnaissance, vulnerability research, scripting support, and evasion of anomaly detection, as well as for development. 
\textbf{(2) CyberAv3ngers}: This group, suspected of being affiliated with the Iranian Islamic Revolutionary Guard Corps (IRGC), employs GPT models to conduct research on programmable logic controllers. CyberAv3ngers is known for disruptive attacks on industrial control systems and programmable logic controllers (PLCs) in sectors such as water, manufacturing, and energy. Their targeted infrastructure is typically associated with Israel, the United States, or Ireland.
\textbf{(3) STORM-0817}: An Iranian threat actor identified as STORM-0817 is developing malware and tools to scrape social media. This actor was found to use OpenAI’s models to debug malware, receive coding assistance for building a basic Instagram scraper, and translate LinkedIn profiles into Persian, aimed at identifying potential targets. 
These three cases ---detected based on credible tips--- illustrate that while AI offers robust defensive capabilities, adversaries are increasingly harnessing these technologies to advance their offensive operations—raising new challenges for global cybersecurity, especially in critical sectors like energy.
\newline
\newline
\textbf{Recommendations.} Building on our findings, we propose:

\begin{enumerate}[leftmargin=*,label=\arabic*)]
  
  \item \textbf{Extension to Other Automation Domains.} To leverage the same generative-AI pipeline to ingest and normalize data from multiple automation sectors. This is applicable to actors, incidents, reports and malware descriptions. While our primary focus is the energy sector, the generative‑AI pipeline and subsequent geopolitical clustering extend readily to other automation environments—such as process industry, automotive manufacturing, or water treatment. This can be easily achieved by swapping the energy‐specific keyword  for a domain‑appropriate one, e.g., using \textit{automotive-related} instead of \textit{energy-related} in the schema definition during prompt engineering (see Figure~\ref{fig:parser_flowchart}). This translates into a simple code update, demonstrating that our method can provide actionable risk insights across diverse automation sectors with minimal reconfiguration.
 The possibility of summarizing results for different domains into an “Automation Threat Dashboard” for real‑time comparative risk assessment is especially interesting.
  \item \textbf{Dynamic Adversarial Augmentation.}  Generating key synthetic  examples of adversarially‑crafted threat descriptions and including them in the prompts to harden the parser against emerging jargon or evasion tactics.
\end{enumerate}

\section{Conclusion}\label{sec:Conclusion}

This paper describes different threat intelligence databases, and introduces a way to parse their unstructured data using generative AI for further analysis. Our findings indicate that cyber threats targeting the energy sector follow distinct patterns compared to general threats. Specialized threat actors frequently emerge as the top origins for energy-focused incidents, coinciding with geopolitical flashpoints. AI-based detection tools—specifically static ML solutions—do not currently outperform traditional antivirus engines for detecting malware used against the energy domain. We observe that AI components are both attack vector and attack surface. These insights, obtained using our generative AI parsing methodology, underscore the need for geopolitical awareness, focused threat intelligence, and improved model prompting tailored to avoid the impact of bias and adversarial perturbations.








\bibliographystyle{unsrtnat}

\end{document}